# CHAPTER X

# Quantum Learning with Noise and Decoherence:
## A Robust Quantum Neural Network

Nam H. Nguyen, Elizabeth C. Behrman, and James E. Steck

## 1. Abstract

Noise and decoherence are two major obstacles to the implementation of large-scale quantum computing. Because of the no-cloning theorem, which says we cannot make an exact copy of an arbitrary quantum state, simple redundancy will not work in a quantum context, and unwanted interactions with the environment can destroy coherence and thus the quantum nature of the computation. Because of the parallel and distributed nature of classical neural networks, they have long been successfully used to deal with incomplete or damaged data. In this work, we show that our model of a quantum neural network (QNN) is similarly robust to noise, and that, in addition, it is robust to decoherence. Moreover, robustness to noise and decoherence is not only maintained but improved as the size of the system is increased. Noise and decoherence may even be of advantage in training, as it helps correct for overfitting. We demonstrate the robustness using entanglement as a means for pattern storage in a qubit array. Our results provide evidence that machine learning approaches can obviate otherwise recalcitrant problems in quantum computing.

## 2. Introduction

Quantum computing may very well be the way to find solutions to a host of calculations that are difficult to do with a classical computer [1]. On the fundamental scale, it offers the opportunity to approach true simulation of physical reality [2]. But when it comes to scaling up from "proof of concept" hardware to practical size, significant problems are yet to be solved. Among the most recalcitrant ones are the issues related to noise [3] and decoherence [4].

The fact that noise is a special problem in quantum computing may seem peculiar. After all, data scientists doing any kind of computation have always known that if there are enough errors in the input, the output will be wrong. One obvious way of guarding against data errors is redundancy: essentially, multiple backups.   But quantum mechanical computers cannot use simple redundancy, because there is no easy way to make copies of unknown quantum states. This is called the "no-cloning theorem," and is an immediate consequence of a quantum system's being in a superposition of states, unknown until measured – and measurement collapses and destroys the state. This is a fundamental rule in quantum mechanics [5]. Classical noise distributions can be handled using the theory of stochastic processes [6], but, again, quantum mechanics limits both the ease and the effectiveness of these kinds of theories.  So other methods need to be used. And in addition, in quantum computation there is also the unique problem of decoherence, which arises from unwanted interactions with the environment. Quantum mechanics is fragile, which is why, on a macroscopic scale, we rarely need to take quantum effects into account: unless the quantum processes are extremely well isolated, the quantum state will decohere and become, essentially, classical. When this occurs in a quantum computer, the quantum nature of the computation is lost. So, if we are specifically interested in doing quantum computing, we need to guard against these kinds of effects as well.

Most researchers who address the problem of noise in quantum computing use the method of ancilla [7,8], which are extra quantum bits (qubits), for error correction. The problem with this approach is the fast growth of the number of ancilla necessary, making scaleup much harder. Some recent papers on the decoherence problem are those of Glickman [9], Takahashi [10], Roszak [11], Dong [12], and Cross [13]. Glickman et al. [9] used the scattering of photons to understand the process of decoherence better. Their experience showed that it might be possible to control decoherence in a quantum system by taking advantage of an atom's spin state. Takahashi et al. [10] were able to use high magnetic fields to suppress quantum decoherence to levels far below the threshold necessary for quantum information processing. Roszak et al. [11] suggested a general approach to protect a two level system against decoherence by engineering a non-classical multiple superposition of coherent states in a non-Markovian reservoir.

Neural network methods are another way of dealing with noise. Because of the distributed nature of the computation and the multiple interconnectivity of the architecture, classical neural networks are inherently robust to noise [14]. Dong et al. [12] presented a systematic numerical methodology called sampling-based control for robust quantum design of a quantum system. This method is similar to ours in that it also uses machine learning. In the training step, the control is learned using a gradient flow based learning algorithm for an augmented system constructed from samples. Cross et al. [13] showed that quantum learning is robust to noise by proving that the complexity of learning between classical and quantum methods is the same. However, when noise is being introduced, the best classical algorithm will have superpolynomial complexity, whereas, the complexity in quantum algorithms is only logarithmic.

Our research group has been investigating the advantages of a machine learning approach to quantum computing for some time [15-17]. Parallel to this we also have been exploring the advantages of a quantum approach to machine learning [15-18]. An exemplar of both thrusts is the calculation of entanglement.

Entanglement, like the no-cloning theorem, is a direct result of superposition. Entanglement is a purely quantum phenomenon not possible with classical states. Mathematically, a quantum bit or qubit can be thought of as a two-state system, in Dirac notation expressible as |0> or |1>, analogous to a classical bit's being either up or down. (This is called the "charge basis".) However a qubit can also be in a complex superposition of both the |0> and the |1> states, not: exclusively up <u>or</u> down and we don't know which, but: both up <u>and</u> down, at the same time. We would express this mathematically as $|\varphi\rangle = a|0\rangle + be^{i\theta}|1\rangle$, where $\{a, b, \theta\}\epsilon\mathbb{R}$. The probability that the qubit $|\varphi\rangle$ would be measured to be in the |0> or the |1> state would be the absolute magnitudes squared of the coefficients, here, $a^2$ or $b^2$, respectively. Now suppose the system consists of two qubits, qubit A and qubit B. A general state of the system could be written as the superposition state $|\psi\rangle = a|00\rangle + be^{i\theta_1}|01\rangle + ce^{i\theta_2}|10\rangle + de^{i\theta_3}|11\rangle$, where $\{a, b, c, d, \theta_1, \theta_2, \theta_3\}\epsilon\mathbb{R}$. This is a state of both qubits but not a product state necessarily: That is, there are states of the joint system that are not expressible as a product state, as (state of A) times (state of B), or $[a_0|0\rangle + a_1|1\rangle] \otimes [b_0|0\rangle + b_1|1\rangle] = a_0b_0|00\rangle + a_0b_1|01\rangle + a_1b_0|10\rangle + a_1b_1|11\rangle$. Consider, for example, the state $|00\rangle + |11\rangle$. If it were a product state, then $a_0b_0 = 1 = a_1b_1$, and $a_0b_1 = 0 = a_1b_0$. These conditions cannot be true simultaneously. The state $|00\rangle + |11\rangle$ is called "entangled", because while A and B do not have definite states, and in fact are equally likely to be either up or down, knowledge (measurement) of either one gives us knowledge of the other: if A is measured to be in the |0> state, then by the Born rule, the system has "collapsed" to the |00> part of the superposition, and B will definitely be measured to be in the |0> state also. Entanglement

is quantum correlation, stronger than anything possible classically. Super-position and entanglement are the source of the power of quantum computation [5].

Since entanglement is a purely quantum phenomenon, its calculation is not possible by any classical computation. Since there is no closed-form solution for the entanglement of a general state of even a two-qubit system, much less of a many-qubit system, there is no quantum algorithm for its calculation, either. The problem seems ideal for a quantum neural network, which could be trained on known exemplars then generalized. Indeed, in previous work [16, 19] we have successfully shown that a quantum neural network can in fact be trained to a general entanglement witness of a two-qubit system. We have also shown [20] that these results can be generalized to three-, four-, and five-qubit systems.

In the present work, we use our machine learned entanglement witness to explore and address the question: Is our quantum neural network robust to noise and to decoherence? Preliminary work on the simple two-qubit system [21] is encouraging. Here, we generalize to larger systems and show that, in fact, the robustness is maintained and even improved with increasing system size. Indeed, it may even be true that the presence of noise improves learning, as it prevents "overfitting" [22]. We also present preliminary results of an application, using entanglement as a means of pattern storage. Techniques of machine learning and neural networks may provide solutions to many of the problems facing large scale quantum computing.

The organization of the rest of this chapter is as follows. In Section 3, we review our method of dynamic learning for a quantum system. In Section 4, we introduce noise and decoherence. Subsection 4.1 then presents results for the 3-qubit system. Subsection 4.2 generalizes our results to 4- and 5-qubit systems. In Subsection 4.3 we calculate the coefficient of determination to show that robustness increases as the size of the system increases. We also briefly discuss the advantage of noise in neural network calculations, that it corrects for overfitting. In Subsection 4.4 we explore alternate methods of adding noise, and show that results are substantially the same. In Section 5 we analyze our results, first, in subsection 5.1, with a stability check; second, in subsection 5.2, with complexity. In Section 6 we present an application: using entanglement between pairs of neighboring qubits as pattern storage. We conclude with a summary in Section 7.

## 3. Dynamic quantum learning

We begin with the general form of the Schrodinger equation:

$$\frac{d\rho}{dt} = \frac{1}{i\hbar}[H, \rho] \qquad (1)$$

where $\rho$ is the density matrix and $H$ is the Hamiltonian. For an N-qubit system, we define the Hamiltonian to be:

$$H = \sum_{\alpha=1}^{N} K_\alpha \sigma_{x\alpha} + \varepsilon_\alpha \sigma_{z\alpha} + \sum_{\alpha \neq \beta=1}^{N} \zeta_{\alpha\beta} \sigma_{z\alpha} \sigma_{z\beta} \qquad (2)$$

where $\{\sigma\}$ are the Pauli operators corresponding to each qubit $\alpha$, $\{K\}$ are the tunneling amplitudes for each qubit to tunnel between states, $\{\varepsilon\}$ are the biases, and $\{\zeta\}$ are the qubit-qubit couplings. We choose the usual charge basis, in which each qubit's state is |0> or |1>.

By introducing the Liouville operator, $L = \frac{1}{\hbar}[..., H]$, equation (1) can be rewritten as:

$$\frac{d\rho}{dt} = -iL\rho \qquad (3)$$

which has the general solution of:

$$\rho(t_f) = e^{-iLt}\rho(t_0) \qquad (4)$$

Notice that the time evolution of the system is a function of the parameters $\{K, \varepsilon, \zeta\}$. That is, if one or more of them is changed, the way a given state evolves in time will also change. Therefore, equation (4) is mathematically isomorphic to the equation for information propagation in a neural network:

$$\phi_i = \sum_j w_{ij} f_j(\phi_j); \quad \phi_{output} = F_W \phi_{input} \qquad (5)$$

where $\phi_{output}$ is the output vector of the network, $\phi_{input}$ is the input vector, and $F_W$ is the network operator, which depends on the neuron connectivity weight matrix W. Thus we can look at our quantum system as a quantum neural network, or QNN. Note that in our QNN, $W \in \mathbb{C}^n$ ; that is, our network takes on complex valued weights. Classically complex valued networks have been shown to be more powerful than real valued networks. For instance, a complex valued neural network can solve the nonlinear XOR problem without a hidden layer [23].

In our quantum neural network computation, the role of the input vector is played by the initial density matrix $\rho(t_0)$, and the role of the output is played by a measurement performed on the density matrix at the final time. For the specific problem of entanglement witness, we used a projective measurement on the z-axis, denoted as $M_z$. Thus the output can be written as $|\text{tr}(M_z \rho(t_f))|$, where tr is the trace. The set of parameters $\{K, \varepsilon, \zeta\}$ play the role of the weights of the network. We then can train the system to evolve in time to a set of particular final states at the final time $t_f$ by adjusting the parameters $\{K, \epsilon, \zeta\}$ through any machine learning approach: supervised, unsupervised, reinforcement, etc. In our work here we use a gradient descent learning algorithm. Time evolution is a quantum mechanical property; hence it can be represented by a unitary operator $U$, which is an isometry, $\|Ux\| = \|x\|$, and surjective on the Hilbert space $\mathcal{H}$. Thus, a quantum mechanical function, such as an entanglement witness of the initial state, can be mapped to an observable (a measure) of the system's final state. This mapping layer is equivalent to applying a non-linear activation function in classical neural network. The detailed derivation and analysis of our learning algorithm for the QNN can be found in [19].

## 4. Learning with Noise and Decoherence

In previous work, we showed that quantum neural computation is robust under random perturbations of the density matrix for the two qubit quantum system [21]. We now generalize this result by extending our previous work to three, four and five qubit quantum systems, and show that the increase in the number of qubits improves robustness to both "noise" and "decoherence." Note that we define "noise" as perturbation to the magnitude of the elements of the density matrix, while "decoherence" is perturbation to the phase. Splitting the density matrix this way is possible since we are working with numerical simulation. Throughout the simulation, in order to conserve probability, the density matrix must remain Hermitian, positive definite, and with unit trace.

As we did with the two-qubit system, we will investigate the effects of noise during both training and testing. Furthermore, we can use knowledge of the smaller system as "partial knowledge" of the larger system, in finding the set of parameter functions that will perform the desired computation. This kind of inference is called, in the literature, "bootstrapping" [24]. In previous work [20], we successfully used this technique to find an entanglement indicator for a general multi-qubit system: we trained the two-qubit system to output an entanglement indicator, then used those functions as the starting point for training the three-qubit system. From three we bootstrapped to four, and so on. In each case, the amount of additional required training decreased, because more and more of the information necessary for the entanglement indicator was present already in the N-1 qubit system.

# 4.1 Results for the three-qubit system: Training and Testing

For a three qubit system, our Hamiltonian is written out explicitly in term of qubits A, B and C as

$$H = K_A \sigma_{xA} + K_B \sigma_{xB} + K_C \sigma_{xC} + \varepsilon_A \sigma_{zA} + \varepsilon_B \sigma_{zB} + \varepsilon_C \sigma_{zC} + \zeta_{AB} \sigma_{za} \sigma_{zB} + \zeta_{BC} \sigma_{zB} \sigma_{zC} + \zeta_{AC} \sigma_{zA} \sigma_{zC} \qquad (6)$$

The $\sigma_{xi}$ or $\sigma_{zi}$ can be written as tensor products of the Pauli matrix $\sigma_x$ or $\sigma_z$ with the 2x2 identity matrix $I$. The order depends on the qubit label. For example, $\sigma_{xB}$ is the outer or tensor product of the identity with $\sigma_x$, $\sigma_{xB} = I \otimes \sigma_x \otimes I$. The identity operators in the first and third places ensure that $\sigma_x$ operates only on the B qubit. The increased connectivity is evident in the increased number of qubit-qubit terms: While with the two-qubit system there is but one connection, with the three-qubit system there are three. Since we are training for a symmetric measure, the tunneling functions are equal: $K_A = K_B = K_C$, and similarly for the $\varepsilon$ and $\zeta$ functions. We now have more output measures {$M_z$} to be trained. We need, however, to be able to distinguish among entanglement between qubits A and B, between A and C, between B and C, and amongst A, B, and C (three-way entanglement; see [20] for details.) Therefore, the number of training pairs, and hence the amount of possible training, goes up like the connectivity; were it not for bootstrapping, this would indeed be a daunting challenge.

For the two-qubit system, we used a training set of four [19]; thus, for the three-qubit system, we need a set of thirteen: three sets of four for the three pairwise entanglements, and one more for three-way entanglement (GHZ state). Table 1 shows our training data for pairwise entanglement when there is no noise added to the system.

| Input state $\lvert \Psi(0) \rangle$ | Target | Trained |
|---|---|---|
| $Bell = \lvert 000 \rangle + \lvert 111 \rangle$ | 1 | 0.998 |
| $Flat = \lvert 000 \rangle + \lvert 010 \rangle + \lvert 100 \rangle + \lvert 111 \rangle$ | 0.0 | $1.3 \times 10^{-5}$ |
| $C = 0.5 \lvert 100 \rangle + \lvert 111 \rangle$ | 0.0 | $1.9 \times 10^{-4}$ |
| $P = \lvert 000 \rangle + \lvert 010 \rangle + \lvert 100 \rangle$ | 0.44 | 0.44 |
| RMS | | $1.8 \times 10^{-3}$ |
| Epoch | | 30 |

Table 1 - The four training pairs with their targets and trained values

The parameter functions $\{K, \epsilon, \zeta\}$ have similar characteristics to the two qubit system parameter functions. In fact, they are identical functions just with different amplitudes. Once these parameters functions are trained, we can use them to test other states. The results on these testing states, which can be either pure or mixed, tell us whether the system has correctly generalized (learned). We found that all three functions $\{K, \epsilon, \zeta\}$ are easily parameterized as simple oscillating functions. Figures 1-3 show the actual trained parameters in dashed lines and their Fourier fit in solid lines.

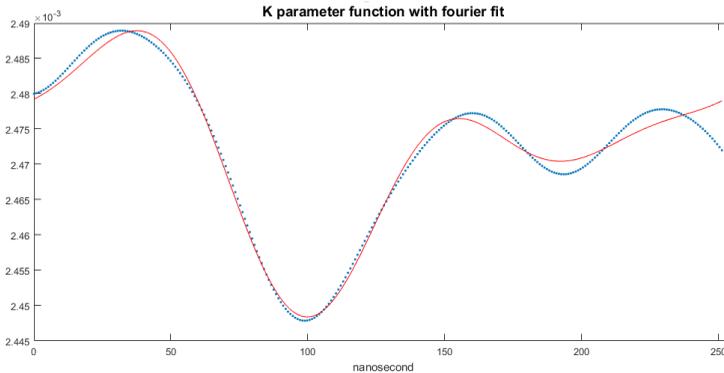

Figure 1: K parameter function trained with zero noise or decoherence, for the 3-qubit system. The solid line is the Fourier fit to the trained values at each time slice (dots.)

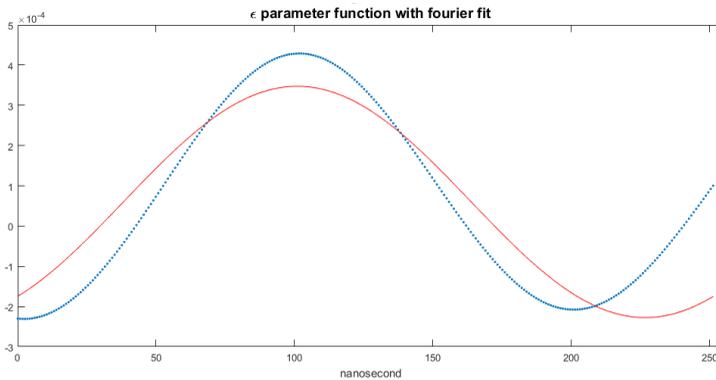

Figure 2: ε parameter function trained with zero noise or decoherence, for the 3-qubit system. The solid line is the Fourier fit to the trained values at each time slice (dots.)

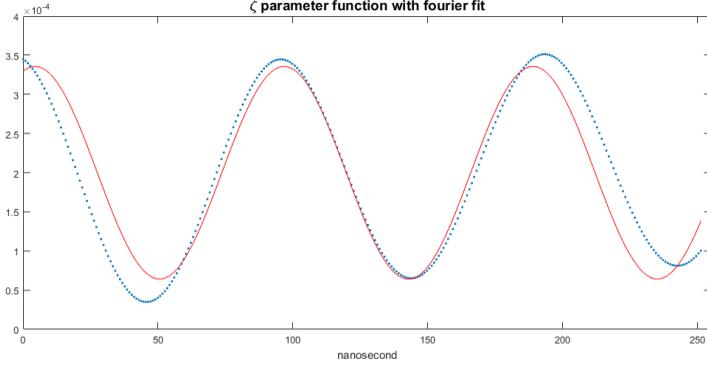

Figure 3: ζ parameter function trained with no noise or decoherence, for the 3-qubit system. The solid line is the Fourier fit to the trained values at each time slice (dots.)

Testing using the Fourier fit parameter functions gives identical (within computational error) results as testing with the trained parameter pointwise data, that is, the differences between the data points and the fits are small enough not to matter. Table 2 shows the fitted parameter functions' Fourier coefficients for zero noise.

$$f(t) = a_0 + a_1 \cos(\omega t) + b_1 \sin(\omega t) + a_2 \cos(2\omega t) + b_2 \sin(2\omega t) + a_3 \cos(3\omega t) + b_3 \sin(3\omega t)$$

| Coefficients | K | epsilon($\epsilon$) | zeta ($\zeta$) |
|---|---|---|---|
| $a_0$ | 0.002472 | $5.9935 \times 10^{-5}$ | $1.9991 \times 10^{-4}$ |
| $a_1$ | $1.071 \times 10^{-6}$ | $-2.3456 \times 10^{-4}$ | $1.2969 \times 10^{-4}$ |
| $b_1$ | $-1.103 \times 10^{-6}$ | $1.6611 \times 10^{-4}$ | $4.0433 \times 10^{-5}$ |
| $a_2$ | $-1.109 \times 10^{-6}$ | --------- | --------- |
| $b_2$ | $1.023 \times 10^{-7}$ | --------- | --------- |
| $\omega$ | 0.02498 | 0.02498 | 0.05750 |
| RMS | $9.774 \times 10^{-7}$ | $1.840 \times 10^{-6}$ | $7.291 \times 10^{-6}$ |

Table 2: Fourier coefficients for 3-qubit fitted parameter functions with no noise and decoherence.

We then proceed by adding noise and decoherence to the system. To carry this through, we first trained the 2-qubit system with no noise or decoherence, then bootstrap [24] the already trained parameter functions {K, $\epsilon$, $\zeta$} to the three qubit system. This helped to decrease the number of epochs needed to train the 3-qubit system [20], since the parameters of the 2-qubit system are similar to the parameters of the 3-qubit system as noted above. We steadily increased the noise perturbation to the density matrix during training, and found that the 3-qubit system parameters are more stable to noise and decoherence than the 2-qubit system. That is, the data points on the parameter functions deviate less as noise, decoherence, or both, are added. Results of the parameter functions for the 3-qubit system, trained at 0.0089 phase noise (decoherence), are shown in Figures 4-6. The level of noise we report is the amplitude, that is, the root-mean-square-average size of these random numbers, imposed at each timestep.

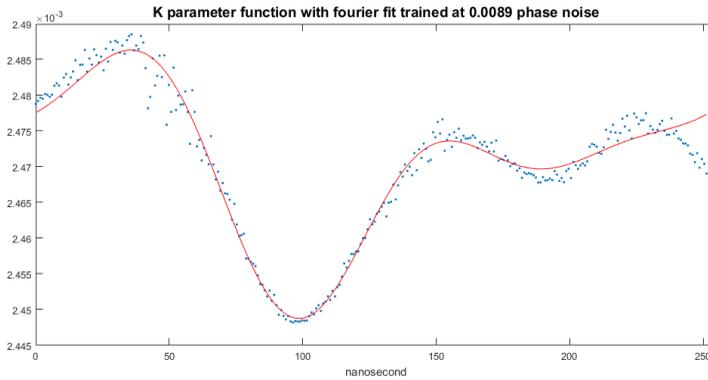

Figure 4: Parameter function K trained at 0.0089 amount of decoherence in a 3-qubit system. The solid red curve represents the Fourier fit of the actual data points.

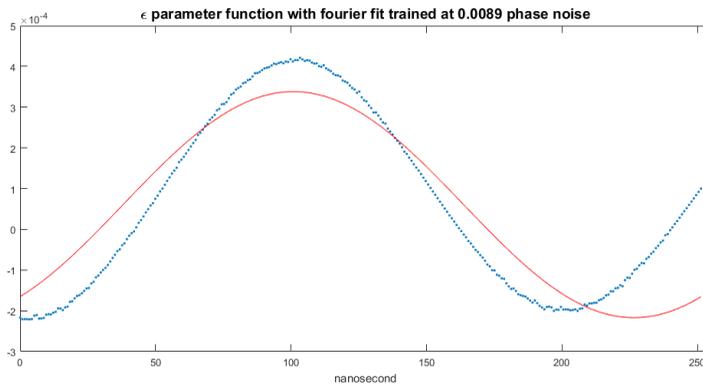

Figure 5: Parameter function ε trained at 0.0089 decoherence in a 3-qubit system. The solid red curve represents the Fourier fit of the actual data points.

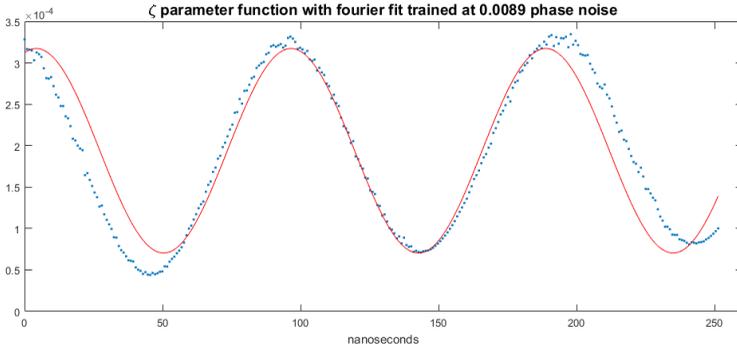

Figure 6: Parameter function ζ trained at 0.0089 decoherence in a 3-qubit system. The solid red curve represents the Fourier fit of the actual data points.

At equal amounts of phase noise added to the density matrix, the data points for the three-qubit system are less scattered than the parameter functions for the two qubit system. Magnitude noise gives similar results. We can also add both noise and decoherence to the system simultaneously (we call this "total noise".) As with the two-qubit system in our previous work [21], we graph the Fourier coefficients of each parameter function as a function of total noise. Figures 7-9 show that the coefficients, and therefore the parameter functions, change very little.

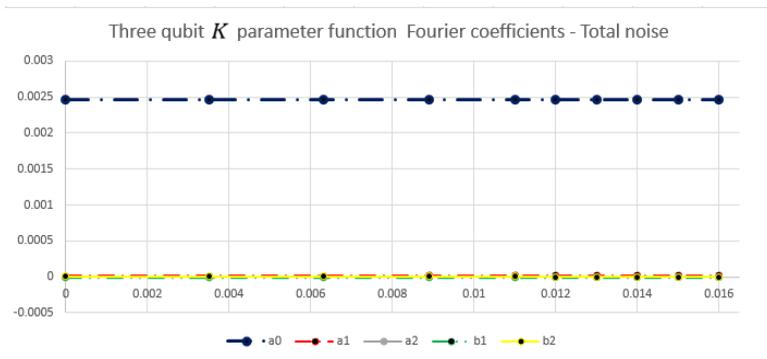

Figure 7: Parameter function K Fourier coefficients as a function of total noise for the 3-qubit system.

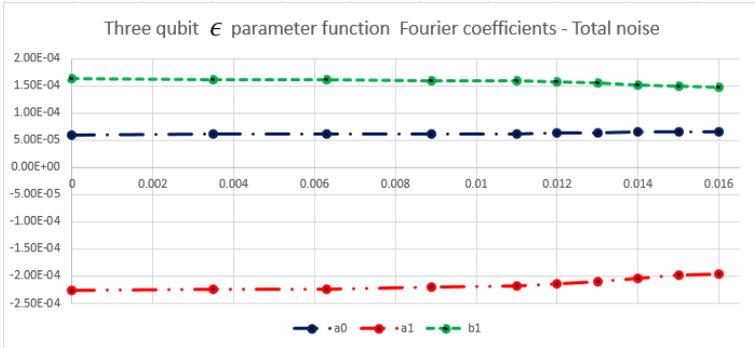

Figure 8: Parameter ε Fourier coefficients as a function of total noise for the 3-qubit system.

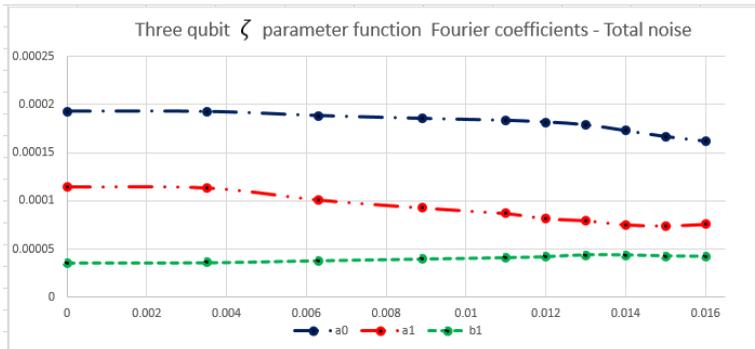

Figure 9: Parameter function ζ Fourier coefficients as a function of total noise for the 3-qubit system.

So far, we have shown that our system is robust to noise and decoherence during the learning process. We need also to test these learned parameters on some "testing" states (i.e., not in the training set) to see how well the system has actually learned, and what is the impact of the noise on the performance. We choose a pure state |Pure> and a mixed state Mix for this process. We choose |Pure> to be $|Pure\rangle = N_P[|000\rangle + \gamma|001\rangle + |011\rangle]$, for $0 \geq \gamma \geq 1$, and where the normalization constant $N_P = [2 + |\gamma|^2]^{-1/2}$ Note that |Pure> is a superposition of the $|000\rangle$ state, the $|001\rangle$ state, and the $|011\rangle$ state. The |Pure> state is pairwise entangled between qubits B and C, decreasingly so as γ increases. Figure 10 shows the testing of this state, as a function of γ, for increasing amounts of (total) noise. There is some spread, but the behavior is qualitatively correct, and the indicator relatively stable.

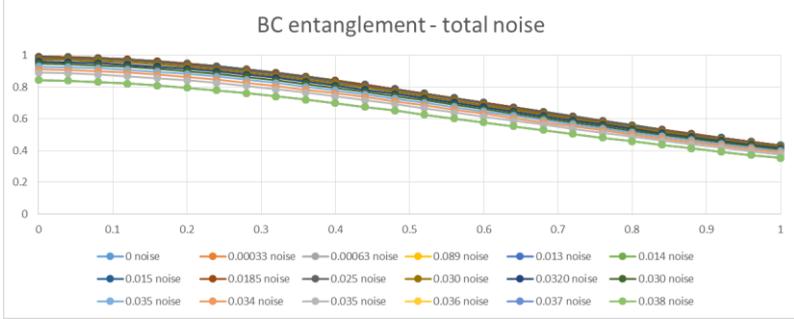

Figure 10: Testing at different noise levels for the pure state |Pure> for a three-qubit system, as a function of γ.

We now choose an exemplar mixed state Mix, whose density matrix ρ_Mix is given by

$$\rho_{Mix} = N_M([|000\rangle + |011\rangle][\langle 000| + \langle 011|] + \gamma |001\rangle\langle 001|) \tag{7}$$

where the normalization constant $N_M = [2 + \gamma]^{-1}$. For this state, we should also expect to have full pairwise entanglement for the BC pair when γ = 0, which should decrease as γ increases. Figure 11 shows testing of Mix at similar noise levels. We notice the same trend as with the previous state: the entanglement indicator is relatively stable.

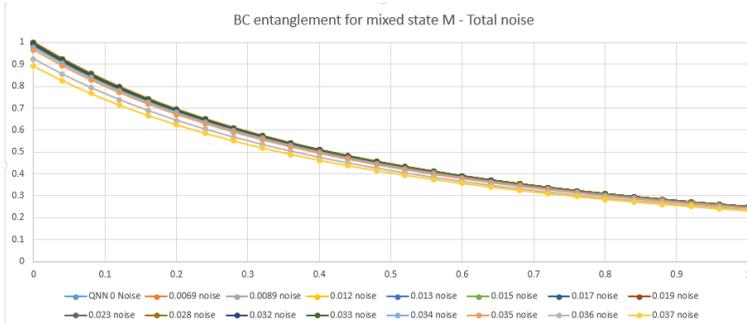

Figure 11: Testing at different noise levels for a mixed state Mix, for the three-qubit system, as a function of γ.

## 4.2 Results for the four- and five-qubit systems: Training and Testing

The evidence of improvement in training and testing as well as the stability to noise and decoherence in a three qubit system is clear. We will now show that the system becomes more stable as we increase its size. This improvement in the results should not be surprising since the bigger the system, the more connectivity we have in our network. Note that for the large N-qubit systems, we train for each level of entanglement: For each pair of qubits, we use the four training states (Bell, P, Flat and C) for that pairwise entanglement, and in addition we also include the training pairs for three-way entanglement, four-way, and so on, up to N-way entanglement (also called GHZ.) See [20] for details. For instance, for the five-qubit system, there are $\binom{5}{2}$ pairs and therefore $\binom{5}{2}$ sets of {Bell, P, Flat, C} states for training the pairwise entanglement; $\binom{5}{3}$ three-way entanglements; $\binom{5}{4}$ four-way entanglements; and $\binom{5}{5} = 1$ five-way GHZ state. As before, since this is a simulation, we can separate the noise and decoherence into two problems, or combine them together to get "total noise". To demonstrate the improvement in robustness, we examine the most noise-sensitive parameter function, K. Figures 12 and 13 show the parameter functions K for the four and five qubit systems respectively.

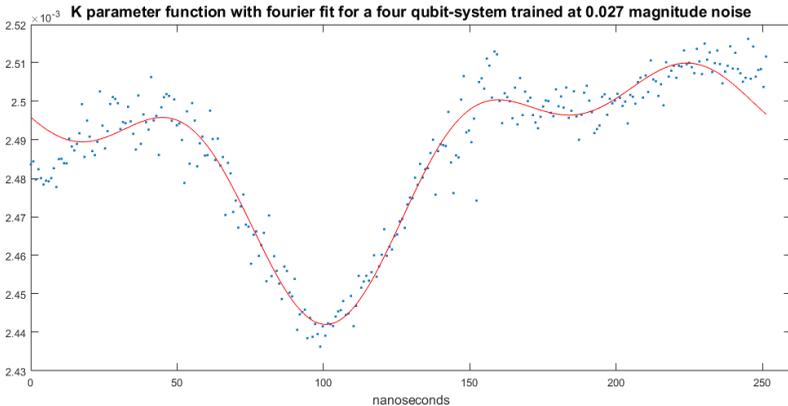

Figure 12: Parameter function K for a 4-qubit system trained at 0.027 level of noise with its Fourier fit.

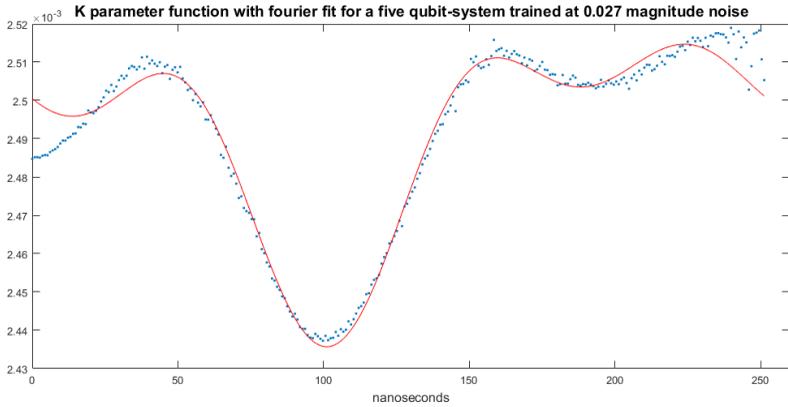

Figure 13: Parameter function K for 5-qubit system trained at 0.027 level of noise with its Fourier fit.

## 4.3 Quantifying the improvement in robustness with increasing size of the system

Note that the noise level being added to the system in Figures 12 and 13 is almost three times as much as the noise added to the three qubit system (Figure 4); these results are, however, comparable. It is obvious there is an improvement in robustness. To quantify that robustness as a function of the number of qubits, we plot the coefficient of determination for each of the least-square curve fits of the trained parameter functions, $R^2$, as a function of the number of qubits in the system, in Figures 14-16. $R^2$ is defined as one minus the (normalized) sum of squares of the residuals; therefore, $R^2$ varies from zero (bad fit) to one (perfect fit).

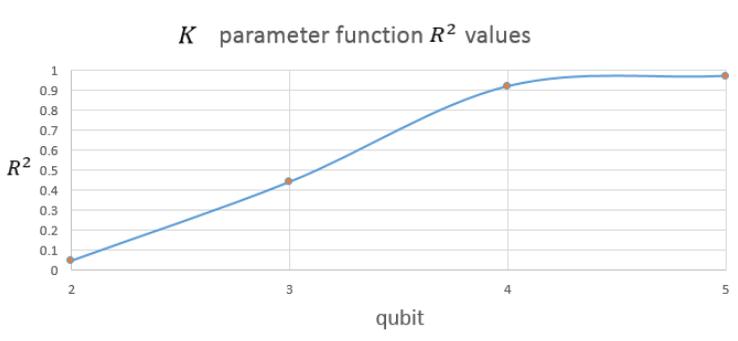

Figure 14: The coefficient of determination, $R^2$, as a function of number of qubits, for the K parameter function, trained at 0.02741 total noise.

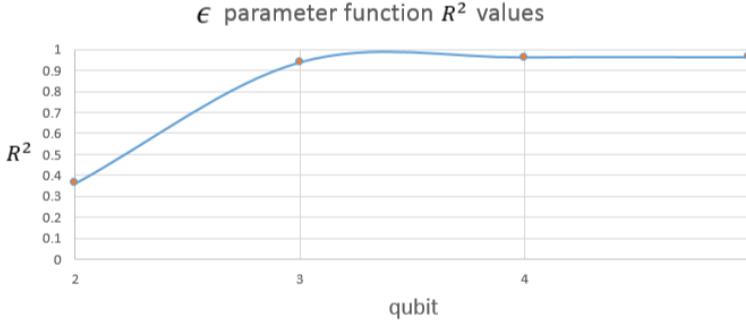

Figure 15: The coefficient of determination, $R^2$, as a function of number of qubits, for the ε parameter function, trained at 0.02741 total noise.

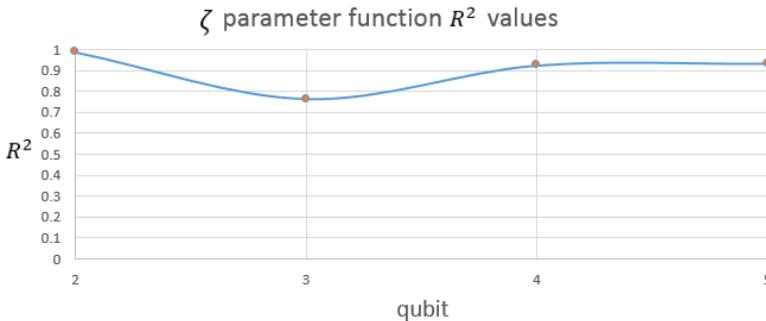

Figure 16: The coefficient of determination, $R^2$, as a function of number of qubits, for the ζ parameter function, trained at 0.02741 total noise.

As the number of qubits increases, $R^2$ increases towards one, and may converge towards an asymptote by four or five qubits. The increase is uniform except for the ζ parameter (Figure 16), but even in that case, $R^2$ never gets below 0.8, and still converges towards a high asymptote for number of qubits of four or five.

This improvement is not unexpected, for two reasons. First, we observed in earlier work that the amount of training necessary decreases sharply as the number of qubits increases [20]; and, second, it is well known that the presence of (small amounts of) noise can in fact improve robustness, because of the problem of "overfitting": If the data set is small, the network essentially memorizes all the training pairs, leading to very bad testing results [22]. This is illustrated with a linear vs high order polynomial fit example, ahown in Figure 17.

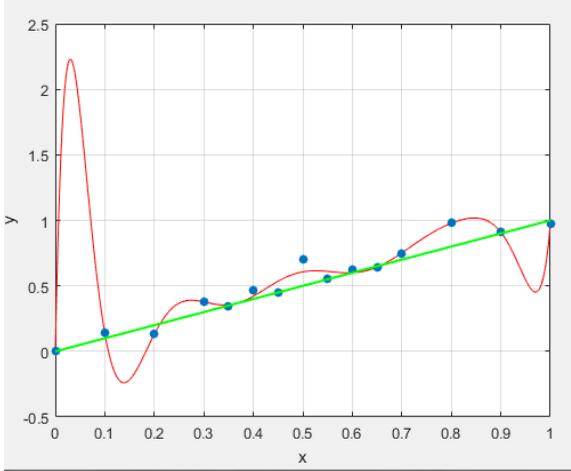

Figure 17: An illustrative example of "overfitting" to the data (blue dots.) The red curve represents a possible function that an overfitting network would have learned comparing to the actual correct fit, the green line. Training error for the red line would be small, but any subsequent testing would give large errors. Adding noise to the data points would correct for overfitting.

Our training sets for the entanglement witness are small compared to the number of free parameters to be determined, that is, the parameter functions {K, ϵ, ζ} as functions of time. Thus the training sets represent only a very small subset of the entire Hilbert space, and overfitting can easily occur. One method to prevent overfitting and to improve the structure of the predicted function in neural networks is to add random noise (usually white Gaussian noise) during training [25-27]. Adding noise will affect the training iteration step but it will spread out the data points and prevent the network from fitting each data points precisely, hence avoiding the overfitting issue and improving the robustness of the network [28]. This is a direct consequence of the inability of the network to memorize the training data since it is continuously being perturbed by noise.

## 4.4 Learning with other types of noise

We have shown that QNN calculation is robust to noise and to decoherence, and that that robustness improves as the size of the system increases. But we need also to address the type of perturbation applied. What we do here, adding white Gaussian noise to the density matrix, is similar to what is known as "random jitter", in which random perturbations are applied to the input data. Noise could instead be added to the outputs and to the gradients of the network during training [27]. Another possibility is to add noise to the activation, which, in our case, would be equivalent to adding noise to our projective measure $M_Z$. Yet another procedure would be to add noise to the weights of the network [26, 29]. This assumes that the parameters themselves are somewhat uncertain or noisy. This is similar to adding noise directly to the Hamiltonian (since the Hamiltonian is itself a function of the parameter functions.) An advantage of this approach is that one need not separately impose conservation of probability, because, as long as the Hamiltonian remains Hermitian, the system will still obey all the physical requirements to be a quantum system. A disadvantage though is that the effects of decoherence cannot be directly investigated. Some preliminary results are shown in Figures 18-21 for the bias (ε) parameter function, trained with magnitude noise being added to the Hamiltonian for the 2-,3-,4-, and 5-qubit systems respectively.

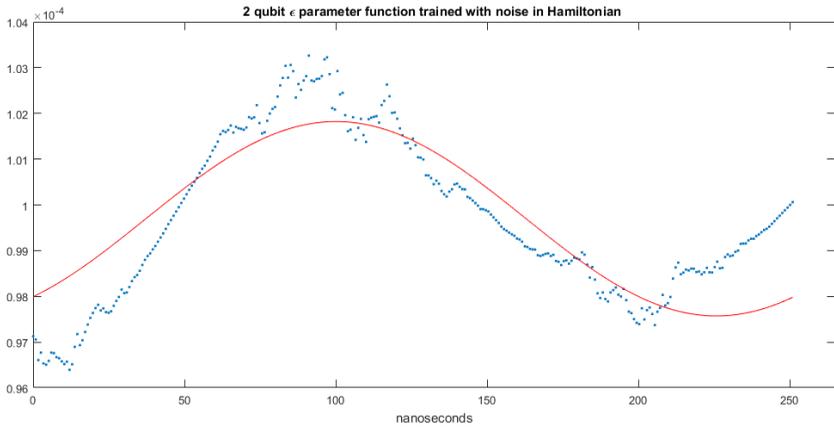

Figure 18: Training the parameter function ε with noise added to the Hamiltonian instead of to the density matrix for a 2-qubit system.

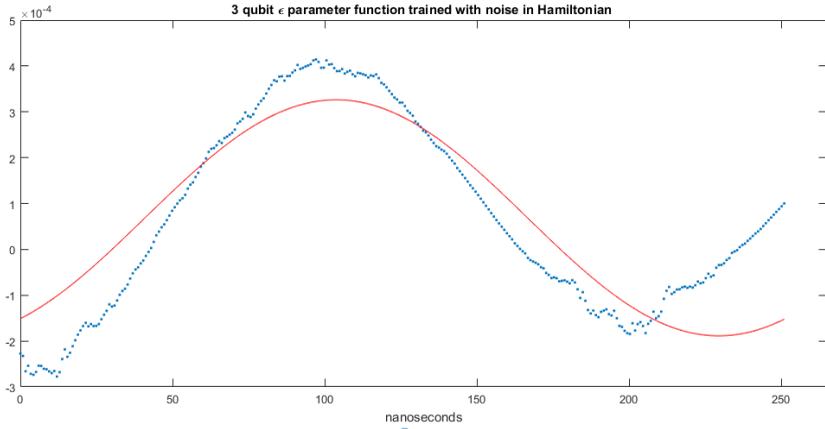

Figure 19: Training the parameter function ε with noise added to the Hamiltonian instead of to the density matrix for a 3-qubit system.

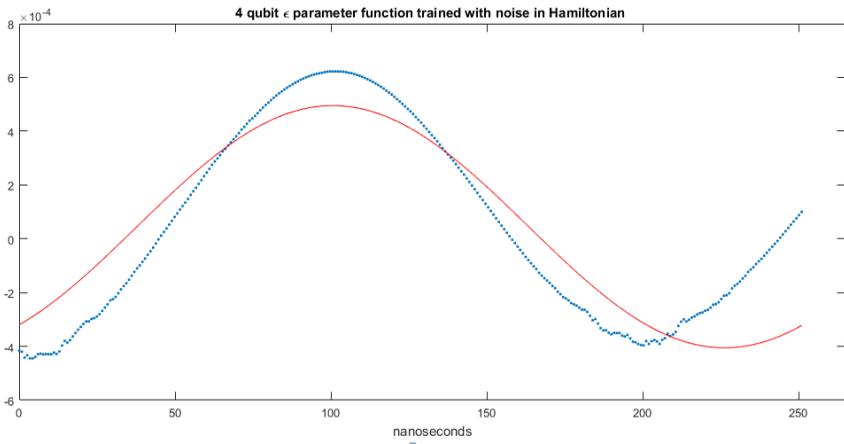

Figure 20: Training the parameter ε with noise added to the Hamiltonian instead of to the density matrix for a 4-qubit system.

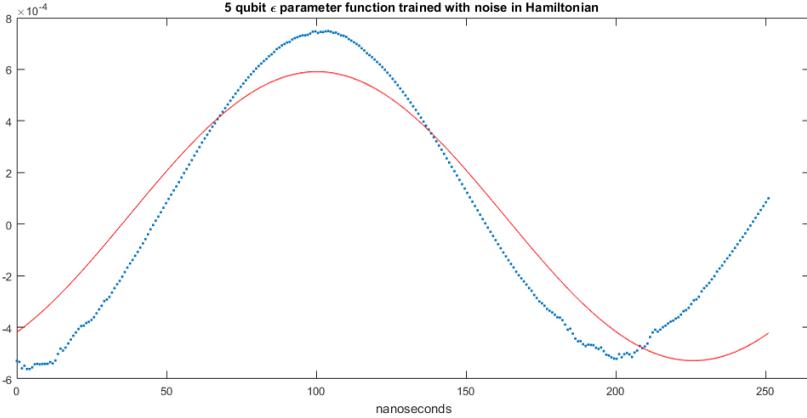

Figure 21: Training the parameter ε with noise added to the Hamiltonian instead of to the density matrix for a 5-qubit system.

The results, shown in Figures 18-21, seem promising: Clearly, robustness increases with system size also with this method of adding noise. Indeed, it would be surprising if this were not true, since the two reasons cited above in Subsection 3.3 would still apply. This is good for practical purposes since, in reality, noise and decoherence can be introduced anywhere during an experiment.

# 5. Analysis of Results

## 5.1 Stability of the calculations

To check the stability of the simulation, we performed an analysis. Our output function for pairwise entanglement is $(\rho(t_0)) = tr\left(M_z\rho(t_f)\right)$, where in this case $M_z = \sigma_z \otimes \sigma_z$, and where $\rho(t_f)$ is given by Equation (4). To add noise, we perturb the original system by adding $\delta\rho$. Our perturbed output then is:

$$O_{pert} = tr\left(M_z(\rho(t_f) + \delta\rho(t_f))\right) = tr\left(M_z\rho(t_f) + M_z\delta\rho(t_f)\right) = tr(A + \delta A) \qquad (8)$$

where we define a matrix $A$ as the product of the projector and the density matrix at the final time, $A \equiv M_z \rho(t_f)$. Let $\lambda_i$ be an eigenvalue of $A$, i.e, $\lambda_i \in \Lambda(A)$. Then by the Bauer-Fike theorem [30], there exists a $\mu_i \in \Lambda(A + \delta A)$ such that

$$|\mu_i - \lambda_i| \leq \kappa(V) \|\delta A\| \qquad (9)$$

where V results from the diagonalization of A, A= $V^{-1}DV$, and $\kappa(V)$ is the conditional number of V. In our case, A is hermitian so V will be unitary, resulting in $\kappa(V) = 1$. For a non-noisy system we have:

$$O = tr(A) = \sum_{i=1}^{n} \lambda_i \qquad (10)$$

where n is the system's size. For a noisy system, we have:

$$O_{pert} = tr(A + \delta A) = \sum_{i=1}^{n} \mu_i \qquad (11)$$

Therefore,

$$\left|O - O_{pert}\right| \leq \sum_{i=1}^{n} |\mu_i - \lambda_i| \leq n \|\delta A\| \qquad (12)$$

Therefore, the difference between the output and the perturbed output is bounded above by the norm of the perturbation of the output matrix $\delta A$. Thus, our simulations are stable: the noise levels are indeed only at the level of a perturbation.

## 5.2 Complexity of the computation

The direct calculation of entanglement for a two-qubit system is relatively simple. Wootters showed [31] that the entanglement of formation could be written in closed form for a general (pure or mixed) state as:

$$E_F = -\frac{1}{2}\left[1 + \sqrt{1 - C^2}\right] \log_2\left(\frac{1}{2}\left[1 + \sqrt{1 - C^2}\right]\right) - \frac{1}{2}\left[1 - \sqrt{1 - C^2}\right] \log_2\left(\frac{1}{2}\left[1 - \sqrt{1 - C^2}\right]\right) \qquad (13)$$

where C is the ``concurrence'', defined for pure states $|\psi\rangle$ by

$$C^2 = \left|\langle\psi|\psi_{sf}\rangle\right|^2 \qquad (14)$$

where the spin flipped state $|\psi_{sf}\rangle$ is

$$|\psi_{sf}\rangle = \sigma_{yA}\sigma_{yB}|\psi^*\rangle \qquad (15)$$

and the asterisk, as usual, indicates the complex conjugate. If we write (wolog) the pure state $|\psi\rangle$ as

$$|\psi\rangle = a|00\rangle + be^{i\theta_1}|01\rangle + ce^{i\theta_2}|10\rangle + de^{i\theta_3}|11\rangle \qquad (16)$$

then the concurrence is given by

$$C^2 = 4[a^2d^2 + b^2c^2 - 2abcd\cos(\theta_3 - \theta_2 - \theta_1)] \quad (17)$$

Therefore, to estimate the entanglement of a 2-qubit pure state using our QNN, our sole requirement is to be able to estimate, or output, three measurements: $a^2d^2 + b^2c^2$, $abcd$, and $\theta_3 - \theta_2 - \theta_1$. (In the special case that all the coefficients are real (in this basis), we only need a single output, and the function is quadratic in the input amplitudes.). This result is also true for any pairwise entanglement of an N-qubit system. Since any admixture only diminishes the entanglement, this gives us a bound for mixed states as well. For three-way entanglements of pure states, we have a closed form solution [32], which is quadratic in the input amplitudes.

Now, we choose as an "output" for the QNN the average value of an experimental measure $M_z$ at the final time $t_f$, which, for a pure state, has the following general form:

$$\langle M_z \rangle = \langle \psi(t_f)|M_z|\psi(t_f)\rangle = \langle \psi(t_0)|e^{iHt/\hbar}M_z e^{-iHt/\hbar}|\psi(t_0)\rangle \qquad (18)$$

For a pure product N-qubit state (the minimum flexibility in QNN training), it is easy to show that each output of this type is a sum of quadratics in the amplitudes of the input state $|\psi(t_0)\rangle$, with linearly independent coefficients, plus sums and products of cosines and sines in each of the phase angles in additional cross terms of the amplitudes. Since each of the parameter functions can be taken to be time varying (as we do here), this essentially means that any single output can have the complexity of almost any reasonably well-behaved function. Therefore, it is not surprising that our QNN can successfully map a smoothly varying function of a quadratic like the pairwise entanglement of formation for pure states. It is somewhat more surprising (and gratifying), that this mapping is relatively easy [19], that it generalizes well to mixed states [19, 20], that it bootstraps well to larger systems with a difficulty that decreases with size [20], and that it is robust to both noise and decoherence [21], with increasing robustness as the system gets larger (the present work). While we have no general closed form solution for N-way entanglement, it is not unreasonable to think that, like pairwise and three-way, N-way is no worse than quadratic in the input ampli-tudes. Indeed, the fact that the QNN readily trains to find N-way entanglement provides experimental evidence that N-way entanglement is in fact no worse than quadratic. For comparison: a standard classical neural network [33] shows comparable errors [34] on either training and testing sets only with a very much larger number of neurons (4 layers with almost a thousand neurons in total), and a comparably larger training set (1000 training pairs, versus the four used here.) The classical neural network also does not generalize from the pure to the mixed state, as the QNN does [34].

## 6. Application: Entanglement as Pattern Storage

Entanglement can be used as a means for pattern storage as follows. We encode the pairwise entanglement between certain qubits to represent the shape of a character. For example, using a four qubit system, we can represent the letter Z as the pairwise entanglement between qubits AB, BC, and CD, as shown in Figure 21.

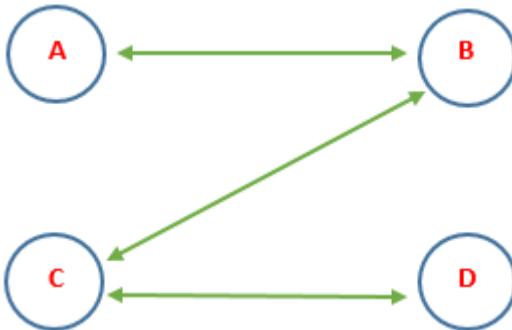

Figure 21: How to encode the letter Z using pairwise entanglement. The green double-headed arrows represent pairwise entanglement.

We can exclude the 3-way and the GHZ state entanglement from our 4-qubit training sets here (since we are only dealing with pairwise entanglement) to help speed up the training process. Since the training and testing is robust to noise to a certain extent, a little corruption in the data will not greatly affect the output from the QNN. Table 3 shows the results of the QNN outputs from different encoded states with noise and decoherence added. Here, we can think of noise and decoherence as corruption to the letters.

With an array of four qubits, we can form only a few characters, but the number of distinct shapes goes up rapidly with the number of qubits: with only six, for example, arranged as a rectangular array in three rows of two, there are 8126 distinct symbols, unique up to translation in the plane. Modern quantum dot technology should allow encoding even of images [35].

These results, while promising, are only preliminary. True "character recognition" will require some means of transferring an image of a letter into a quantum state before performing this process. We are currently extending this work to address this problem, and to include a more complete analysis [36].

| | QNN OUTPUT | | | |
|---|---|---|---|---|
| Magnitude noise and Decoherence | Letter Z | Letter N | Letter O | Letter X |
| 0.0050 | Correct | Correct | Correct | Correct |
| 0.0125 | Correct | Correct | Correct | Correct |
| 0.0250 | Correct | Correct | Correct | Correct |
| 0.0375 | Correct | Correct | Correct | Correct |
| 0.0500 | Correct | Correct | Correct | Correct |
| 0.0625 | Correct | Correct | Incorrect | Correct |
| 0.0750 | Incorrect | Incorrect | Incorrect | Correct |

Table 3: QNN output to encoded states of different letters, showing robustness to noise.

# 7. Conclusions

We have successfully shown that our model of QNN calculation is robust to both noise and decoherence up to five qubits. The increased connectivity both decreases the required training time per qubit and increases the stability of the system, independently of the details of the noise source. This seems to be especially true for decoherence: The increased robustness is evident whether the system is trained or tested with noise. While exact simulations on macroscopic quantum computers remain impractical, these results brings us a step forward in the investigation of the quantum neural approach for extrapolation to macroscopic quantum computing.

A useful quantum computer will have thousands of qubits. Based on our work so far, it seems likely that the increase in size will only improve our training and help our system to be more robust to noise and decoherence. We chose an entanglement indicator as an example problem, but we would expect the same kind of effects doing other measures, for the reasons outlined above.

As the number of qubits increases, the number of epochs required for training each qubit decreases with the method of bootstrapping, but the total simulation time necessary goes up. This is to be expected since the reduction is linear but the connectivity is quadratic. Because of this, our calculations were limited to a five-qubit system. However, we have built up to seven-qubit system training only the $\binom{N}{2}$ pairs of each of the Bell, P, Flat and C states. We can see similar results in that case as well. With better hardware, there would be no difficulty in extending our simulations to a much larger system using our bootstrapping technique. In an upcoming paper [37] we validate our results using the Microsoft [38] and IBM [39] encoders; once macroscopic quantum computers are available, we can implement "online" training, and simulations will no longer be necessary.

## Acknowledgements

We thank Mohammed Moustafa and Henry Elliott for helpful discussions and for providing the NeuralWorks data and the six-qubit array symbol data, respectively.